\documentclass[twocolumn,floatfix,amssymb,aps,prb]{revtex4-2}
\usepackage{graphicx}
\usepackage{float}
\usepackage{amsmath}
\usepackage{pifont}
\newcommand{\newpar}{\noindent}
\expandafter\ifx\csname package@font\endcsname\relax\else
 \expandafter\expandafter
 \expandafter\usepackage
 \expandafter\expandafter
 \expandafter{\csname package@font\endcsname}%
\fi
\DeclareRobustCommand\substyle{\name@idx{document substyle}}%
\DeclareRobustCommand\classoption{\name@idx{document class option}}%
\DeclareRobustCommand\classname{\name@idx{document class}}%
\def\name@idx#1#2{%
 {\ttfamily#2}%
 \index{#2\space#1=\string\ttt{#2}\space#1}\index{#1>#2=\string\ttt{#2}}%
}%

\begin{document}
\title{Quantum critical fans from critical lines at zero temperature}
\author{Hui Yu and Sudip Chakravarty}%
\affiliation{Mani L.Bhaumik Institute for Theoretical Physics \\Department of Physics and Astronomy, University of California Los Angeles, Los Angeles, California 90095, USA}


\begin{abstract}
Quantum critical phenomena influences the finite temperature behavior of condensed matter systems through  quantum critical fans whose extents are  determined by the  exponents of the zero temperature criticality.  Here we  emphasize the aspects of quantum critical lines, as discussed previously, and study an exactly solved model involving a transverse field Ising model with added three-spin interaction. This model has three critical lines. We compute the spin-spin correlation function and extract the correlation length, and   identify the crossovers: quantum critical to quantum disordered, or renormalized classical regimes. We construct the quantum critical fans along one of the critical lines. In addition, we also construct finite temperature dynamic structure factors. We hope this model will become experimentally realizable in the future, and our results could stimulate  studies in many  similar models  
\end{abstract}
\maketitle

\section{INTRODUCTION}
 Quantum critical point (QCP) \cite{ref1,ref2,ref3,ref4,ref5} is a point in the parameter space where a continuous phase transition takes place at zero temperature. A significant part of research in condensed matter physics  is focused on describing various quantum phases  and transitions between them. Thus, QCP has become a widely studied subject. Exactly at the critical point, we have quantum fluctuations taking place at all length scales. It is  interesting  to   probe these fluctuations and  the implied quantum behavior of the ground states. However, in reality, all experiments are carried out at finite temperatures and it is necessary to learn how such ground state properties can be deduced from finite temperature measurements. This is accomplished by measuring the finite temperature correlation lengths, dynamic structure factors, and other physical properties. The QCP  leaves fingerprints at nonzero temperatures of these properties. Its influence can be felt in a broad regime, called the quantum critical fan, whose extent depends on the quantum critical  exponents. This idea was successfully exploited in \cite{ref6,ref7} in the context of two-dimensional quantum antiferromagnets. 
 
As an example, suppose we have a cusp-like quantum critical fan with $T_x\sim|g-g_{c}|^{z\nu}$, where $g$ is a tuning parameter, $z$ is a dynamical critical point, and $\nu$ is a critical exponent, and $T_x$ is the crossover temperature between two different quantum phases. If  $z\nu$ is very large, this will make the quantum critical fan very narrow and thus limit the ability to probe quantum fluctuations at finite temperatures. 
On the other hand, If we have quantum critical points with smaller values of $z\nu$, experimental evidence of quantum criticality could be more easily observed at finite temperatures.

The one-dimensional transverse field Ising model (TFIM) is a classic example of QCP. Theoretically, the integrability of the model gives us the power to study its properties in detail. A complete discussion on that topic can be found in \cite{ref4}. Experimentally, this model is well-captured by 
$\mathrm{CoNb_{2}O_{6}}$ \cite{ref8}, which illustrates the nature of quantum criticality. 

Here, as in our recent paper, we consider an exactly solved model which has three interesting {\em critical lines} that goes beyond the notion of a {\em critical point}. It's a three-spin extension of the more familiar Ising model in a transverse field, TFIM. The model is  solved by  Jordan-Wigner and Bogoliubov transformations. The corresponding phase diagram was introduced by Kopp and Chakravarty~\cite{ref9}. Later on, the critical lines in this model were studied and their  topological aspects were discussed by Niu {\em et al} \cite{ref10}. Different phases were identified by the number of Majorana modes on each end of an open chain. Subsequently, we calculated the momentum, $k$, and frequency, $\omega$, dependent dynamical structure factor $S(k,\omega)$ in pure and disordered versions of the model at {\em zero temperature} in a recent paper \cite{ref11}. However, the influence of temperature on the phase diagram remained unexplored, especially from the perspective of the quantum critical fans. This is what we aim  in this paper. It will be important in understanding  experimental observations, if and when such a model is realized and studied in practice.

The  paper is organized as follows. In Sec. II, we introduce our model and the phase diagram, and discuss a few of the properties in the context of quantum critical lines. In Sec. III, we discuss three regimes that appear in finite temperatures and review how correlation length behaves in each such regime. In Sec. IV, we  discuss the method for calculating the correlation function using Pfaffian method. In Sec. V we  discuss our results. The final Section, Sec. VI, is a summary.

\section{The Model and its phase diagram}

The model we consider here is a 3-spin extension of the TFIM, studied previously by \cite{ref9,ref10,ref11}. We first discuss its phase diagram and several characteristics of its critical lines. We will focus on topics that were not studied previously. The Hamiltonian, $H$, is
\begin{equation}
H=-\sum_i(h_i \sigma_i^x+\lambda_2^{\prime}\sigma_i^{x}\sigma_{i-1}^z\sigma_{i+1}^z+\lambda_1^{\prime}\sigma_i^z\sigma_{i-1}^z)
\end{equation}
$\sigma^{x}$ and $\sigma^{z}$ are the standard Pauli matrices.  
Presently, we shall set $h_i= h= cst$. The Hamiltonian after Jordan-Wigner transformation \cite{ref12,ref13} is
\begin{align}
   \sigma_i^x = 1-2c_i^\dagger c_i 
\end{align}
\begin{eqnarray}
\sigma_i^z=-\prod_{j<i}(1-2c_j^\dagger c_j)(c_i +c_i^\dagger)
\label{Eq:JW-2}
\end{eqnarray}
is
\begin{equation}
\label{eq:Ham1}
\begin{split}
H&=-\sum_{i=1}^{N}h (1-2c_i^\dagger
c_i)-\lambda_1^{\prime}\sum_{i=1}^{N-1}(c_i^\dagger c_{i+1}+c_i^\dagger
c_{i+1}^\dagger+h.c.)\\&-\lambda_2^{\prime}\sum_{i=2}^{N-1}(c_{i-1}^\dagger
c_{i+1}+c_{i+1}c_{i-1}+h.c.).
\end{split}
\end{equation}
In contrast to the spin model, the spinless fermion Hamiltonian is actually a one-dimensional {\em mean-field} model of a $p$-wave  superconductor \cite{ref14}, when there are both nearest- and next-nearest neighbor hopping, as well as condensates---note the pair creation and destruction operators. The solution of the corresponding spin Hamiltonian through Jordan-Wigner transformation is, however, exact and includes all possible fluctuation effects and is {\it not a mean-field solution of any kind}.

\begin{figure}[htb]
\includegraphics[scale=0.425]{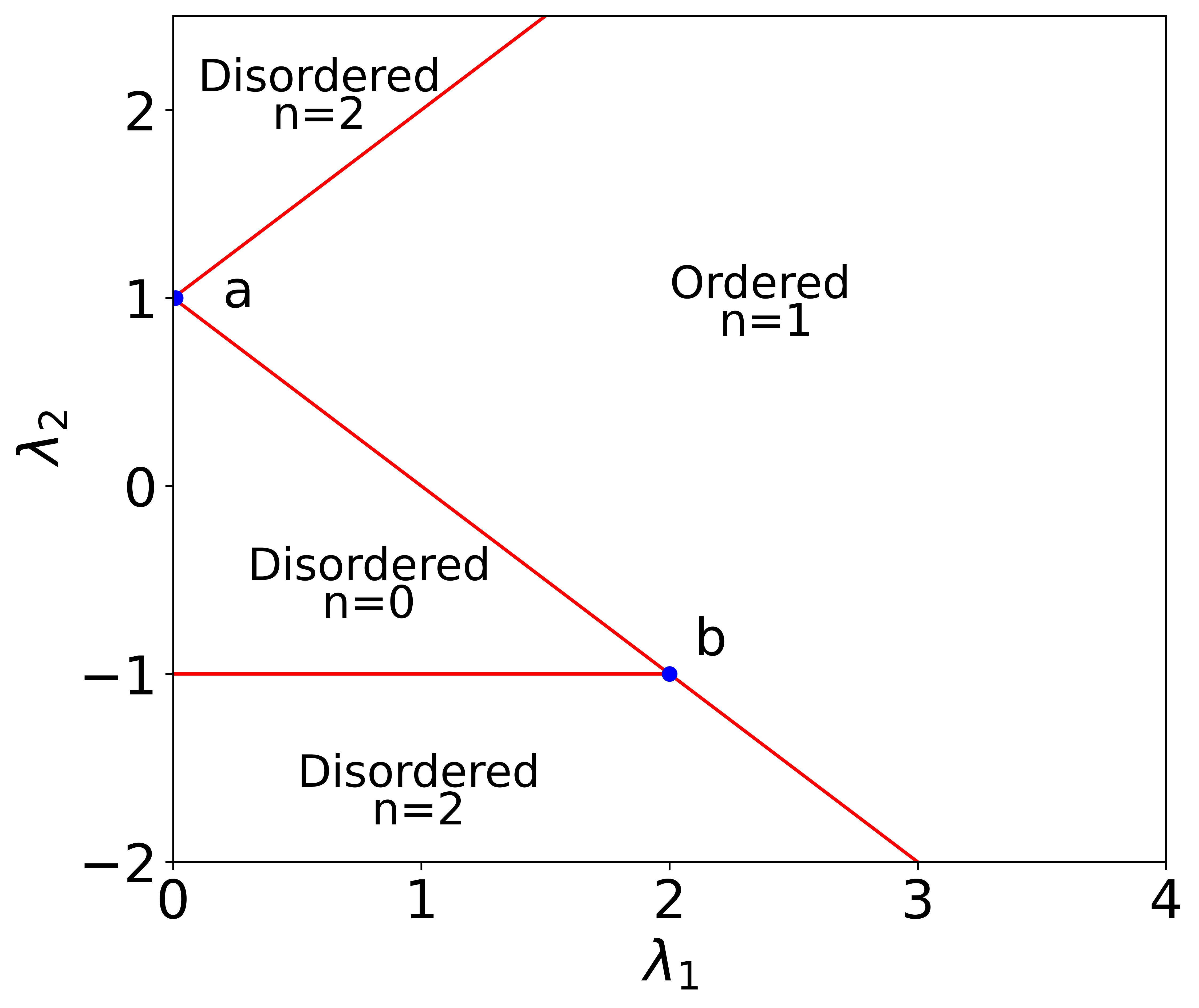}
\caption{The phase diagram: $n =$ 0,1,2, corresponds to regions with $n$ Majorana modes at each end of an open chain. Three quantum critical lines $\lambda_{2}=\lambda_{1}+1$, $\lambda_{2}=1-\lambda_{1}$ and $\lambda_{2}=-1$($0<\lambda_{1}<2$). These lines denote the collapse of the energy gaps of the energy gaps. Points $a$ and $b$ are multicritical points with dynamical critical exponent $z=1$ and $z=2$ respectively.}
\label{fig:QCP}
\end{figure}

Imposing periodic boundary condition, the Hamiltonian can be diagnolized by a Bogoliubov transformation
\begin{equation}
H=\sum_{k} 2E_{k}\left(\eta_{k}^{\dagger}\eta_{k}-\frac{1}{2}\right).
\label{eq:diagonalH}
\end{equation}
As usual, the anticommuting fermion operators $\eta_{k}$'s are suitable linear combinations in the momentum space of the original Jordan-Wigner fermion operators. The spectra of excitations are (lattice spacing will be set to unity  throughout the paper unless stated otherwise)
\begin{equation}
E_{k}=h\sqrt{1+\lambda_{1}^{2}+\lambda_{2}^{2}+2\lambda_{1}(1-\lambda_{2})\cos k -2\lambda_{2}\cos 2k}
\end{equation}
Here $\lambda_1=\lambda_1^{\prime}/h$ and $\lambda_2=\lambda_2^{\prime}/h$ are the scaled coupling constants. 
Quantum phase transitions of this model are given by the nonanalyticities of the ground state energy:
\begin{equation}
E_{0}=-\sum_{k}E_{k}. 
\end{equation}
The derivative of the ground state energy vanishes at $k=0,\pm\pi$ and $\cos k= \lambda_1(1-\lambda_2)/4 \lambda_2$. The nonanalyticites are defined by the critical lines where the energy gaps collapse.  The phase diagram can also be understood from the Majorana zero modes, which we explain below. For the time being refer to Fig.~\ref{fig:QCP}.

\begin{enumerate}
\item For TIFM  without a three-spin interaction,
the gaps collapse at the Brillouin zone boundaries, $k=\pm \pi$ at the self-dual point $\lambda_{1}=1$ and $\lambda_{2}=0$.
\item As we move along the critical line $\lambda_{2}=1-\lambda_{1}$, there are no additional critical points until we reach a multicritical point $\lambda_{2}=1$, where  the gaps collapse at $k=0$. At exactly $\lambda_{1}=0$ and $\lambda_{2}=1$, we have the dynamical critical exponent $z=1$ due to the linearly vanishing spectrum at $k=0$.  Then $\lambda_{2}=1+\lambda_{1}$ constitutes a critical line with criticality at $k=0$.
\item Moreover, the gaps also
collapse at incommensurate points $k=\cos^{-1}(\lambda_{1}/2)$ for  $\lambda_{2}=-1$ and $0<\lambda_{1}<2$. This constitutes an unusual incommensurate critical line. Right at $\lambda_{1}=2$ and $\lambda_{2}=-1$,  we have a non-Lorentz invariant multicritical point with dynamical critical exponent $z=2$.  The spectra vanish quadratically at $\pm \pi$ due to  confluence of two Dirac points.
\end{enumerate}
In the spin representation, our model exhibits two phases - ordered and disordered. These phases are distinguished by the presence of long-range order. As shown in Fig,~\ref{fig:Correlation} the long-range order is reflected in the equal-time correlation function $C(r,0)$ from Eq. 19. Both $\lambda_{2}=1+\lambda_{1}$ and $\lambda_{2}=1-\lambda_{1}$ separate these phases. However, the line $\lambda_{2}=-1$ ($0<\lambda_{1}<2$) can not be understood from symmetry breaking quantum phase transition since it separates two quantum disordered phases. 

 In the fermion language, the phase transitions are best described by the number of Majorana zero modes, $n$, at each end of an open chain, which can be determined numerically. This  was discussed in great detail in a previous paper \cite{ref10}. So $\lambda_{2}=-1$ ($0<\lambda_{1}<2$) is a line of topological transition seperating $n=0$ and $n=2$ Majorana zero modes at each end of the chain. The number of Majorana modes are also winding numbers \cite{ref15,ref16} explained in terms of Anderson pseudospin Hamilitonian \cite{ref17}, when the time-reversal symmetry is preserved. More recently, the topological nature of the model was explained  from the notion of a curvature renormalization group \cite{ref18,ref19}, which  may be useful in higher-dimensional systems.

Finally, this model has a dual representation in which it is equivalent to an anistropic $XY$-model with a magnetic field in the $z$-direction. It is possible that the $XY$-version  is better realized in experimental systems. This dual representation is defined by the dual spin operators $\mu_x$, $\mu_y$, and 
$\mu_z$ such that
\begin{eqnarray}
\mu_x(n) &=& \sigma_z(n+1)\sigma_z(n),\\
\mu_z(n)&=& \prod_{m\le n} \sigma_x(m),
\end{eqnarray}
which implies that 
\begin{eqnarray}
[\mu_z(n),\mu_x(n)]&=& -2i\mu_y(n),\\
\mu_y(n)&=&-i \bigg(\prod_{m \le n}\sigma_x(m)\bigg) \sigma_z(n+1)\sigma_z(n).
\end{eqnarray}
The Hamiltonian under  duality transforms to
\begin{equation}
\begin{split}
H_D&=-\frac{2}{1+r}\sum_n\bigg[\frac{1+r}{2}\mu_x(n)\mu_x(n+1)\\&+\frac{1-r}{2}
\mu_y(n)\mu_y(n+1)+h_{z}\mu_z(n))\bigg],\label{Hxy}
\end{split}
\end{equation}
where we have carried out the rotations : $\mu_x(n)\to \mu_z(n)$, $\mu_z(n)\to \mu_x(n)$, $\mu_y(n)\to -\mu_y(n)$.  The parameters are related  by
\begin{equation}
\lambda_1 = \frac{2h_{z}}{1+r}, \;
\lambda_2 = \frac{r-1}{1+r}.
\end{equation}
The critical line in the $XY$-model, separating the disordered phase from the ordered phase, is at $h_{z}=1$, which corresponds to $\lambda_1+\lambda_2=1$,
separating the ordered phase from the disordered phase. Since the ordered and the disordered phases are exchanged under duality, the disordered phase of
the three-spin model is $\lambda_1+\lambda_2<1$.  

\section{Finite Temperature Crossovers}
\begin{figure}[htbp]
\begin{center}
\includegraphics[scale=0.45]{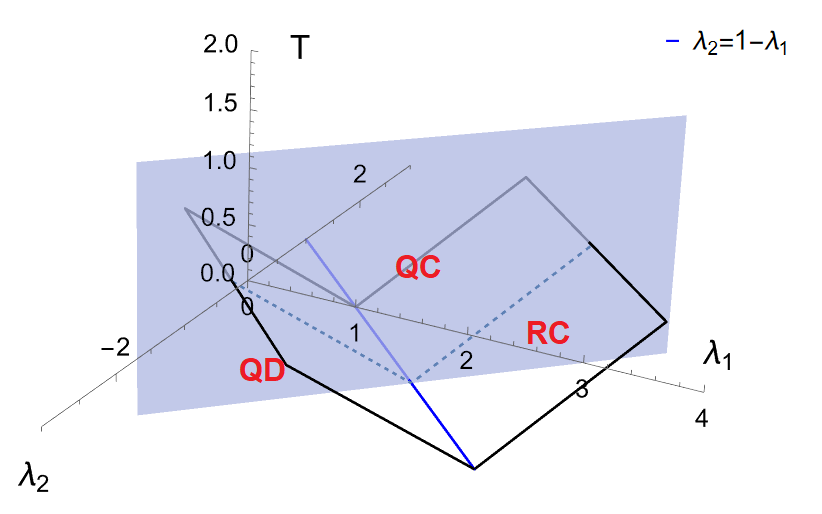}
\caption{ A sketch of a quantum critical fan of a quantum critical line $\lambda_{2} = 1- \lambda_{1}$. The black solid lines are quantum critical fans. The blue dashed lines are the usual quantum critical fan from a critical point which is a point of intersection between the critical line and the blue dashed line. QC: quantum critical. RC: renormalized classical. QD: quantum disordered. }
\label{fig:QCFan}
\end{center}
\end{figure}

In this section, we discuss  quantum critical fans and their crossovers at finite temperatures. Generically, as we raise the temperature, the effect of quantum criticality from a QCP can be felt in an extended region (quantum critical fan) of the parameter space. The width and the shape of the quantum critical fan depend on the critical exponents $\nu$ and $z$.  
In the two-dimensional parameter space, one can approach a critical point in any direction. This leads to a cone-like quantum critical fan for a critical point. In our model, with added three-spin interaction, we have a quantum critical line made out of a line of critical points. Referring to Fig.~\ref{fig:QCFan}, the quantum critical fan looks like a valley along the critical line in this case. The  blue planes denote  the fans. These fans are crossover lines that separate different regimes.  These regimes can be distinguished by the temperature dependence of the correlation length $\xi$ and the relative magnitudes of the   energy scales.   
The following delineates the regimes pertaining to quantum criticality;
\begin{enumerate}
\item Quantum Critical ($\Delta_{gap}\ll T$):
In this regime, the physical properties of the model at finite temperatures are completely determined by the quantum critical point at zero temperature. Tthe correlation length behaves as a power law in $T$.
\begin{equation}
\xi \sim \frac{1}{T^{1/z}}
\end{equation}
where $z$ is the dynamical critical exponent of the QCP.
\item Renormalized Classical ($\Delta_{gap}\ll T$) :
In this regime, $\xi$ goes to $\infty$ exponentially fast as $T$ goes to zero due to the presence of long-ranged  order at zero temperature. In general, we expect the correlation length to have the following form.
\begin{equation}
\xi \sim C_{1}(T)e^{C_{2}/T}
\end{equation}
where $C_{2}$ is a positive constant and $C_{1}(T)$ is a function of $T$. The exact form of $C_{1}(T)$ is not important to us since we are only interested in the general form of $\xi$.
\item Quantum Disordered ($\Delta_{gap}\gg T$):
Since there is no long-ranged order at $T=0$, we expect the correlation length to become temperature independent as $T$ goes to aero,  saturating to a value of order unity.
\begin{equation}
\xi \sim {\rm Const.}
\end{equation}

\end{enumerate}
There is one quantum disordered regime called the oscillatory disordered regime that we will come back to later.

\section{Finite temperature Correlation function}
The signature of the quantum criticality may be discovered by obtaining the correlation length in the neutron scattering experiments \cite{ref20,ref21}. For this purpose, we compute spin-spin correlation function. Here we  discuss the method for calculating the correlation function and the correlation length. One can consult our previous paper \cite{ref11} if one is interested in a full-detailed derivation. 

\begin{figure}
\begin{center}
\includegraphics[scale=0.43]{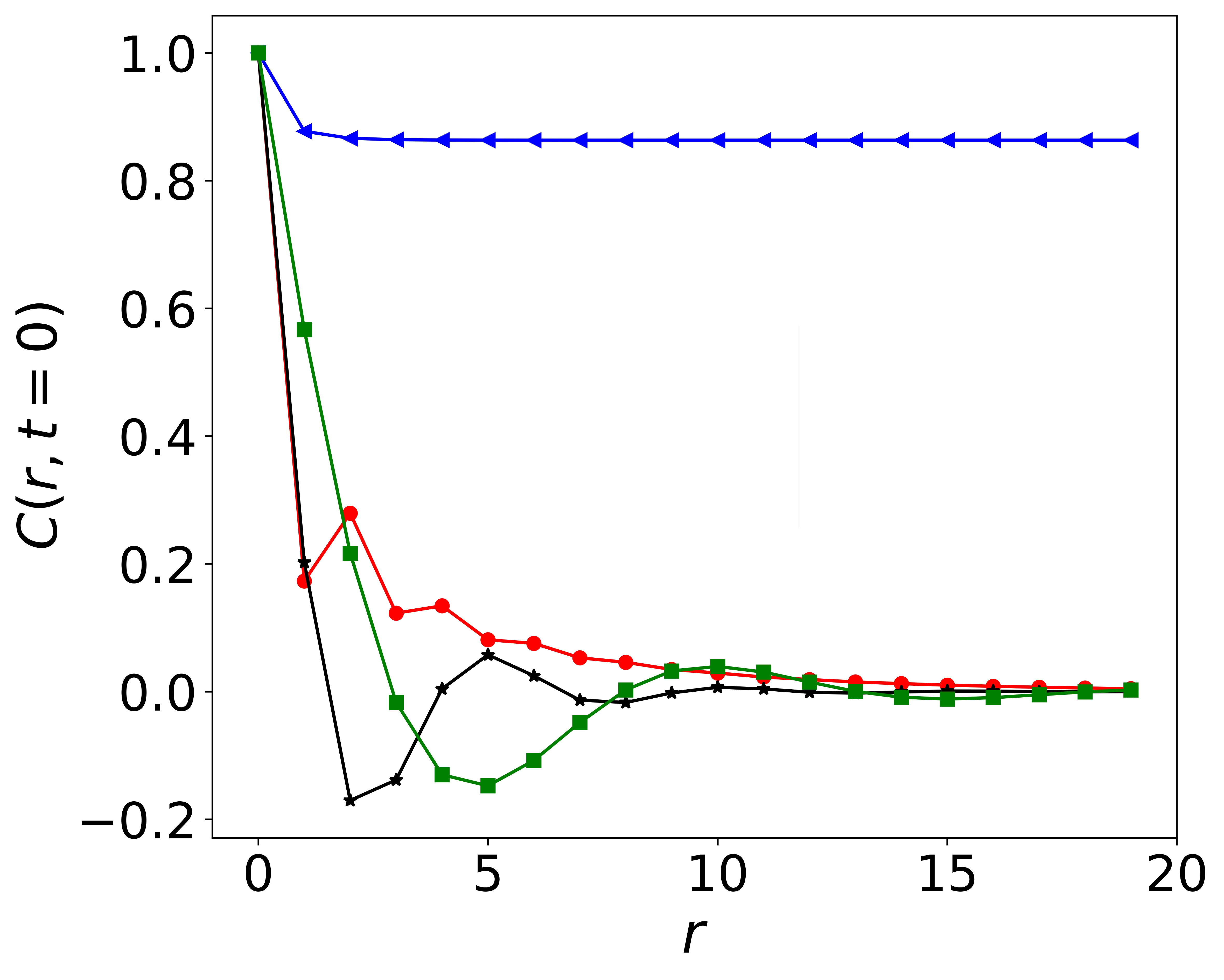}
\caption{Representative equal-time correlation function $C(r,0)$ at $T=0$ in various regions which are labelled by the number of Majorana zero modes $n$ in Fig.\ref{fig:QCP}. The distance $r$ is expressed in  units of lattice spacing. The calculation of the correlation function follows from the Pfaffian representation discussed in detail below. Blue: ($\lambda_1 = 1.5$, $\lambda_2=0$), Red: (${\lambda}_1 = 0.5$, ${\lambda}_2 = 2$).
Black: ($\lambda_1 = 0.5$, $\lambda_2= -0.5$). Green: ($\lambda_1 = 2$, $\lambda_2 = -1.5$.)}
\label{fig:Correlation}
\end{center}
\end{figure}

Quite generally, the spin-spin correlation function $C_{ij}(t)$ is defined as 
\begin{equation}
C(r,t)=\left<\sigma_{i}^{z}(t)\sigma_{j}^{z}(0)\right>
\end{equation}
where $i$, $j$ are lattice sites and $r$ is the separation between them. And the angular  brackets represent a thermodynamic average $\langle(...)\rangle = {\rm Tr}(e^{-\beta H}(...))/{\rm Tr}(e^{-\beta H})$. The equal-time correlation function is
\begin{equation}
C(r,0)=\left<\sigma_{i}^{z}\sigma_{j}^{z}\right>
\end{equation}
In a finite system of length $L$, we choose $i$, $j$ in the middle of the chain to reduce the boundary effects. 
Using the Jordan-Wigner transformation, Eq.~\ref{Eq:JW-2},
we get 
\begin{widetext}
\begin{equation}
C(r,0)=\left<\Biggl(\prod_{m=1}^{i-1}(c_{m}^{\dagger}+c_{m})(c_{m}^{\dagger}-c_{m})\Biggl)(c_{i}^{\dagger}+c_{i})\Biggl(\prod_{l=1}^{j-1}(c_{l}^{\dagger}+c_{l})(c_{l}^{\dagger}-c_{l})\Biggl)(c_{j}^{\dagger}+c_{j})\right>
\label{Eq:Correlation}
\end{equation}
\end{widetext}
Because of the free fermion nature of the Jordan-Wigner transformed Hamiltonian, we can apply Wick's theorem \cite{ref22} to $C(r,0)$. After collecting all terms in Wick expansion, we get a Pfaffian:
\begin{equation}
C(r,0) =   Pf(S) 
\end{equation}
Here $S$ is a $2(i+j-1)$ dimensional skew-symmetric matrix. If we identify $A_{m} = c_{m}^{\dagger}+c_{m}$ and $B_{n} = c_{n}^{\dagger}-c_{n}$, The matrix $S$ is 
\begin{widetext}
\begin{equation}
S =
\begin{pmatrix}
0 & <A_{1}B_{1}> & <A_{1}A_{2}> & <A_{1}B_{2}> & ... & <A_{1}A_{j}> \\
-<A_{1}B_{1}> & 0 & <B_{1}A_{2}> & <B_{1}B_{2}> & ... & <B_{1}A_{j}>\\
-<A_{1}A_{2}> & -<B_{1}A_{2}> & 0 & <A_{2}B_{2}> & ... &
<A_{2}A_{j}>\\
\vdots & \vdots & \vdots & \vdots & \ddots & \vdots \\
-<A_{1}A_{j}> & -<B_{1}A_{j}>  & -<A_{2}A_{j}> & -<B_{2}A_{j}> & ... & 0
\end{pmatrix}
\end{equation}
\end{widetext}
All we need to compute is the two-point correlation function such as 
\begin{equation}
\left<[c_{m}^{\dagger}\pm c_{m}](c_{l}^{\dagger}\pm c_{l})\right>.
\end{equation}
This can be done by utilizing the free fermion operators $\eta_{\mu}$ and $\eta_{\mu}^{\dagger}$ from Eq.\ref{eq:diagonalH}. The results are the following
\begin{equation}
<A_{i}A_{j}> = \sum_{p=1}^{L}\phi_{pi}\phi_{pj}
\end{equation}
\begin{equation}
<A_{i}B_{j}> = \sum_{p=1}^{L}\phi_{pi}\psi_{pj}tanh(\beta E_{p})   
\end{equation}
\begin{equation}
<B_{i}A_{j}> = -\sum_{p=1}^{L}\psi_{pi}\phi_{pj}\tanh(\beta E_{p})   
\end{equation}
\begin{equation}
<B_{i}B_{j}> = -\sum_{p=1}^{L}\psi_{pi}\psi_{pj}  
\end{equation}

\begin{figure}[htbp]
\begin{center}
\includegraphics[scale=0.35]{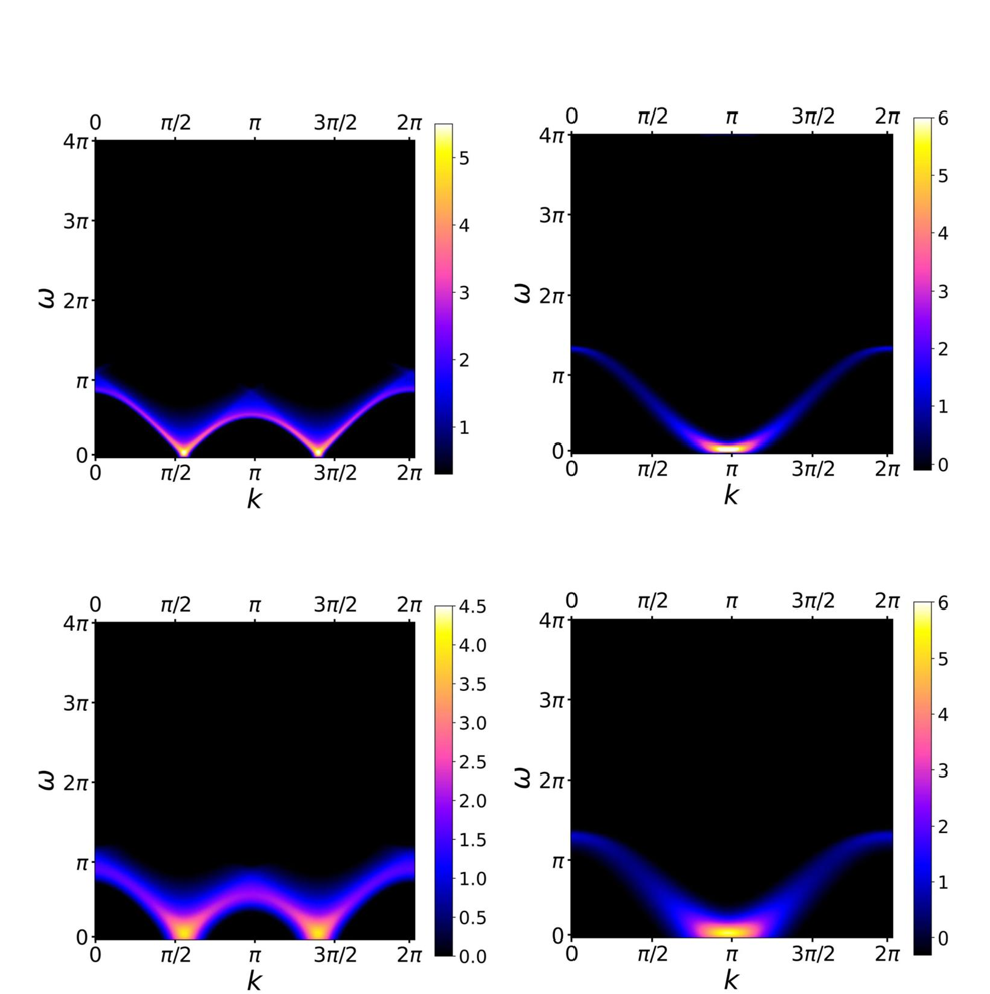}
\caption{Finite temperature $S(k,\omega)$. Thermal brodening is evident. Top Left: ($\lambda_1 = 0.5$, $\lambda_2 = -1$,  $T=0.01$). Bottom Left: ($\lambda_1 = 0.5$, $\lambda_2 = -1$,  $T=0.1$). Top Right: ($\lambda_1 = 2$, $\lambda_2 = -1$,  $T=0.01$). Bottom Top: ($\lambda_1 = 2$, $\lambda_2 = -1$,  $T=0.1$).}
\label{fig:Finite S(k,w)}
\end{center}
\end{figure}
\noindent where $\beta$ is $1/T$. Three $L\times L$ matrices $\phi$,$\psi$ and $E$ come from singular value decomposition (SVD) \cite{ref23} and $M$  obtained by rewriting $H$ from Eq. 4 into a matrix with $c^{\dagger}+c$ and $c^{\dagger}-c$ as its basis. Here $c^{\dagger}+c \equiv (c_{1}^{\dagger}+c_{1},c_{2}^{\dagger}+c_{2},...,c_{L}^{\dagger}+c_{L})$.
\begin{equation}
M=\phi E \psi^{T} 
\end{equation}
\begin{equation}
H =\begin{pmatrix}c^{\dagger}+c & c-c^{\dagger}\end{pmatrix}
   \begin{pmatrix}
0 & M^{T} \\
M & 0
\end{pmatrix} 
\begin{pmatrix} c^{\dagger}+c \\c^{\dagger}-c  \end{pmatrix}
\end{equation}
\noindent
The exact form of $M$ is
\begin{equation}
M = \begin{bmatrix}
    h & -\lambda_{1} & -\lambda_{2} &  &  \\
     & h & -\lambda_{1} & -\lambda_{2}  & & \\
     &  & h & -\lambda_{1} & \ddots & \\
     &  & & \ddots & \ddots & -\lambda_{2} \\
     &  & & & \ddots & -\lambda_{1} \\
     &  & & &   & h 
\end{bmatrix}
\end{equation}
the exact diagonalization of $H$ is not numerically stable (suffers from large errors) if the eigenvalues of $H$ are close to $0$. Thus, instead of directly diagonalizing our $2L\times2L$ Hamiltonian, we chose to use SVD to diagonalize a $L\times L$ matrix.

We still need to deal with one last step. The computation of a Pfaffian consumes a lot of time by standard methods for a large-size system. One of the authors in collaboration  invented an efficient method for dealing with such Pfaffian in \cite{ref24}.  Let $X$ be a $2N \times2 N$ skew-symmetric matrix which has the following form
\begin{equation}
X=\left[ {\begin{array}{cc}
   A & B \\
   -B^{T} & C \\
  \end{array} } \right]
\end{equation} 
\newpar
where $A$ is a $2\times 2$ matrix, and $B$ and $C$ are matrices of appropriate dimensions.Then we have the identity
\begin{equation}
\left[ {\begin{array}{cc}
   I_{2} & 0 \\
   B^{T}A^{-1} & I_{2N-2} \\
  \end{array} } \right]X\left[ {\begin{array}{cc}
   I_{2} & -A^{-1}B \\
   0 & I_{2N-2} \\
  \end{array} } \right]
  =\left[ {\begin{array}{cc}
   A & 0 \\
   0 & C+B^{T}A^{-1}B\\ 
  \end{array} } \right]
\end{equation}   
\newpar
where $I_{n}$ is a $n\times n$ identity matrix, and
\begin{equation}
\mathrm{det}(X)=\mathrm{det}(A)\mathrm{det}(C+B^{T}A^{-1}B)
\end{equation}
\newpar
This gives us an iterative method. We will get a $2\times2$ matrix $A$ at each iteration step; then we treat $C+B^{T}A^{-1}B$ to be our next $X$ and keep doing this. Our $\det(X)$ eventually becomes a product chain of $2\times2$ matrices.

Generically, the equal-time correlation function $C(r,0)$ decays exponentially at finite temperatures. This allows us to determine the correlation length $\xi$ by fitting $C(r,0)$ to an exponential function.
\begin{equation}
C(r,0) \sim e^{-r/\xi}
\end{equation}
where the prefactor could be a constant or an oscillatory function of $r$, as shown below in Sec. V.

\section{Computational Results}
\subsection{T=0, correlation function}
First, we provide some results for $C(r,0)$ at $T=0$. The most remarkable result is the oscillatory quantum disordered phase.
A complex calculation~\cite{ref25} of the instantaneous spin-spin correlation function
showed  that within the ferromagnetic phase in the {\bf dual} representation, Eq.~\ref{Hxy}, there is an oscillatory   phase in  which the 
connected correlation function has oscillatory decay.  The oscillatory phase in the $XY$-model
is bounded by $r^2+h^2\leq 1$, which corresponds to 
$\lambda_2\leq-\lambda_1^2/4$ in the three-spin model. Some of the representative $C(r,0)$ are shown in Fig.~\ref{fig:Correlation}.

\begin{figure}[htbp]
\begin{center}
\includegraphics[scale=0.335]{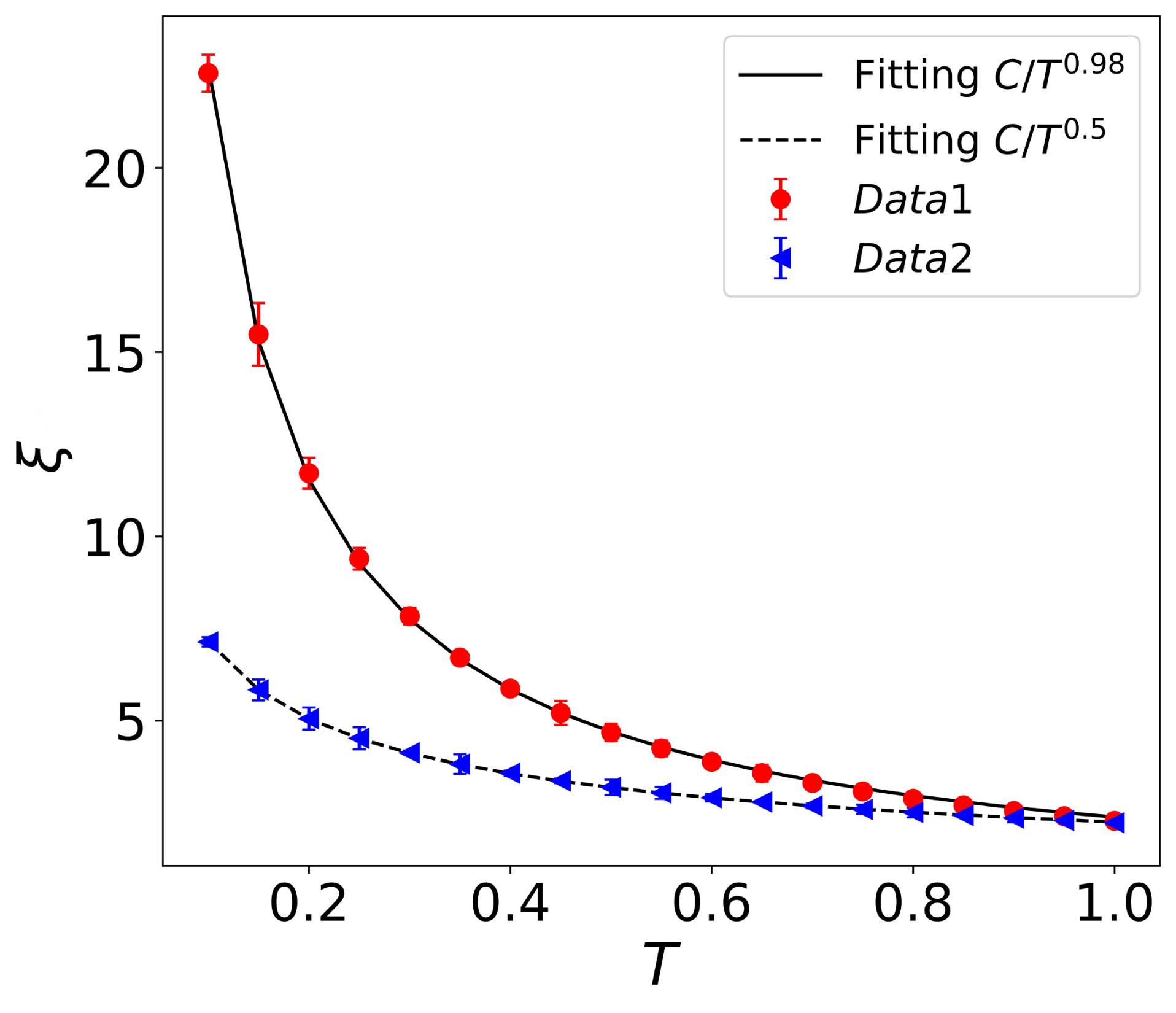}
\caption{The correlation length $\xi$ vs temperature $T$ at two quantum critical points $\lambda_{1}=1$, $\lambda_{2}=0$ ($z=1$) and $\lambda_{1}=2$, $\lambda_{2}=-1$ ($z=2$). $C_{top}\sim 2.38$ and $C_{bottom}\sim 2.25$. }
\label{fig:critical}
\end{center}
\end{figure}

\subsection{Finite temperature $S(k,\omega)$}
We now turn to  discussion of the finite temperature results. This requires unequal time correlation functions, $C(r,t)$.\cite{ref11} The calculations in the following sections were performed on a chain that has 300 lattice sites with free boundary conditions at a number of temperatures; we just show only the results at two different temperatures. The temperature $T$ is measured in units of $h$. We show our finite temperature dynamical structure factor at two critical points $\lambda_1 = 0.5$, $\lambda_2 = -1$ and $\lambda_1 = 2$, $\lambda_2 = -1$ in Fig.~\ref{fig:Finite S(k,w)}.

\begin{figure}[htbp]
\includegraphics[scale=0.43]{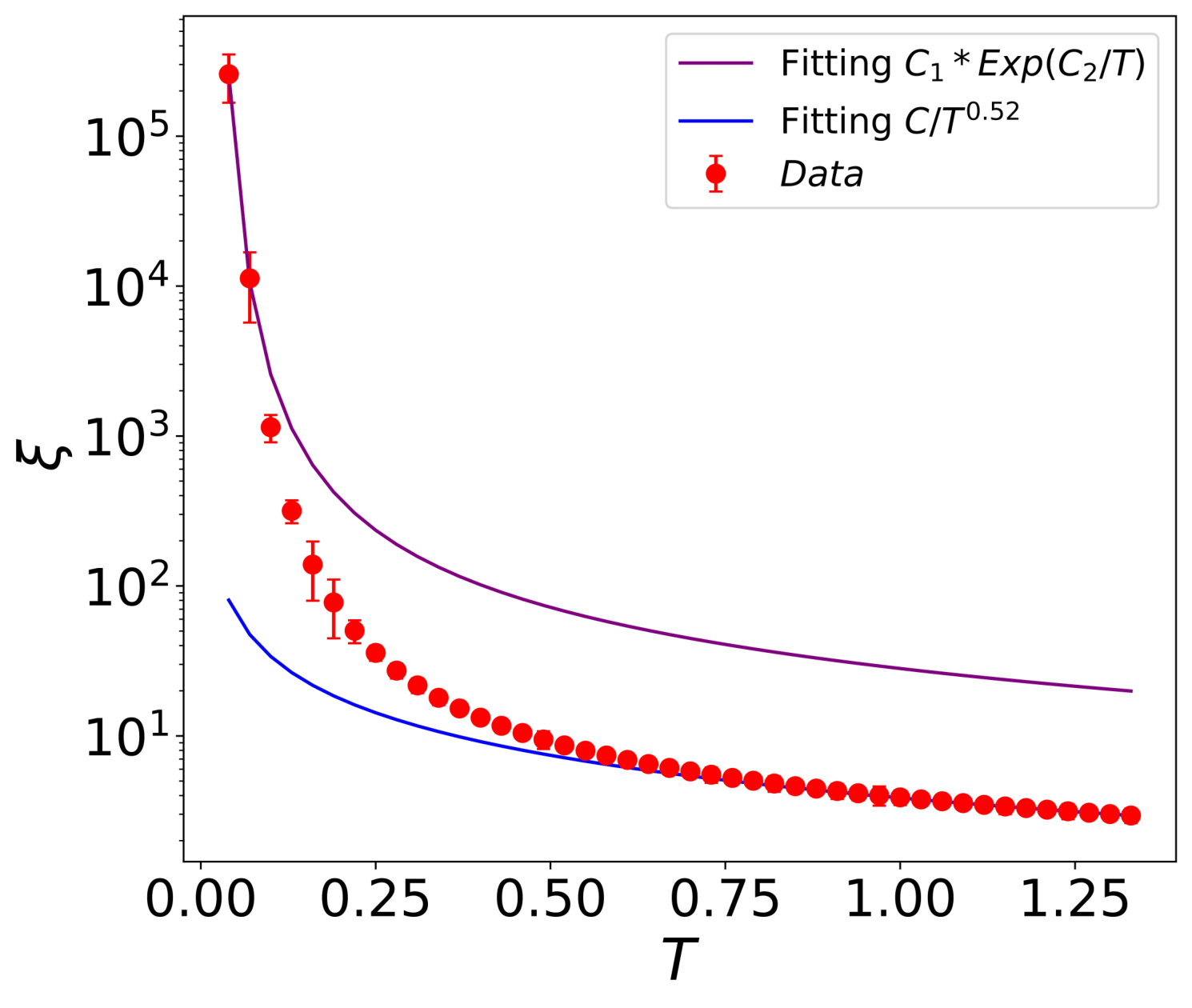}
\caption{Quantum Critical to Renormalized Classical. The correlation length $\xi_{CL}$ vs temperature $T$ at $\lambda_{1}=2.075$, $\lambda_{2}=-0.925$ ($\Delta_{gap}=0.3$). $C\sim 3.81$,$C_{1}\sim 22$ and $C_{2}\sim 0.25$.}
\label{fig:QC TO RC slice}
\end{figure}

\begin{figure}[htbp]
\includegraphics[scale=0.16]{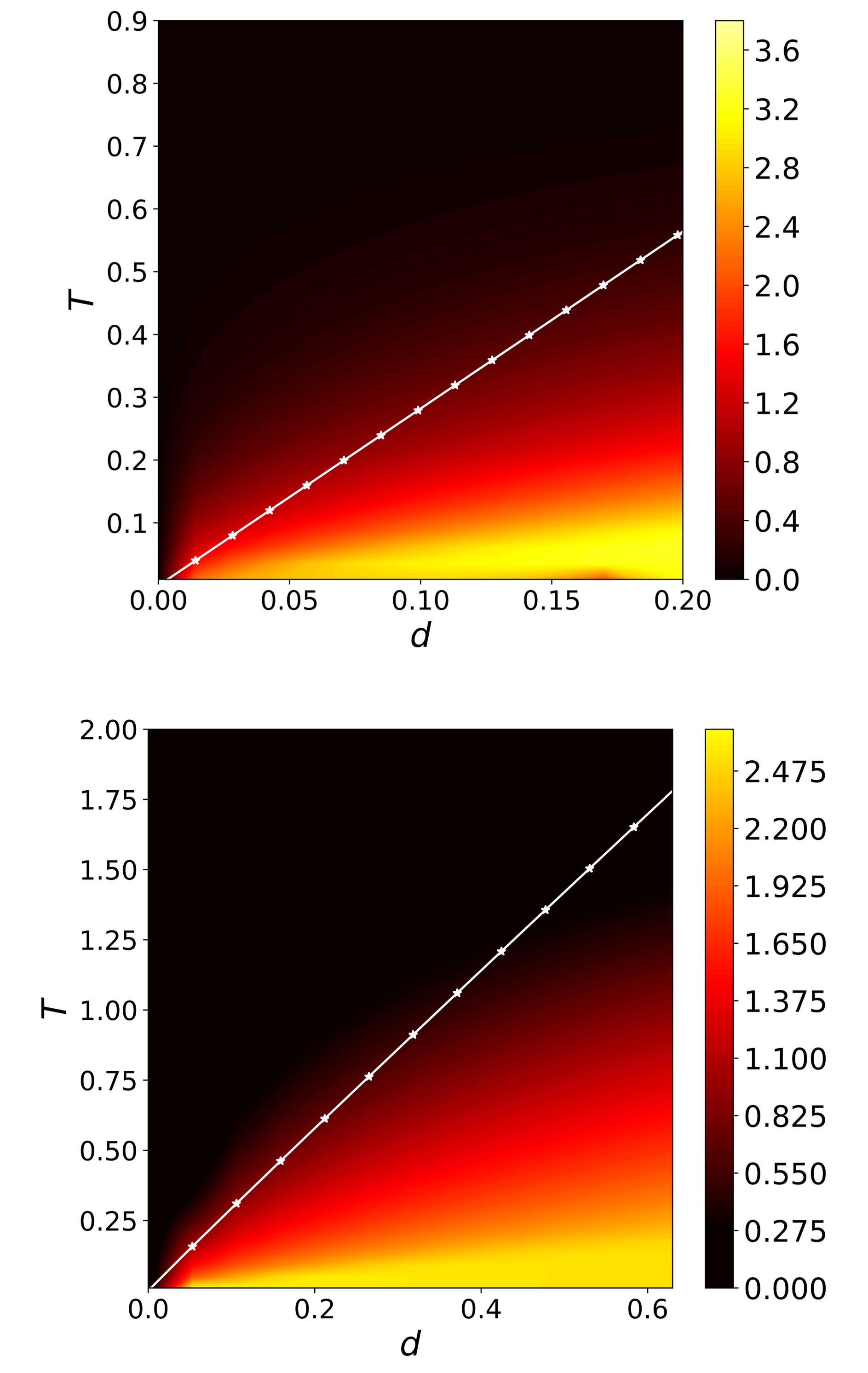}
\caption{Quantum critical fan diagrams. Black: Quantum Critical. Yellow: Renormalized Classical. Top: $d$ is the distance to the multicritical point $\lambda_{1}=1$,$\lambda_{2}=0$ ($z=1$). Bottom: $d$ is the distance to the multicritical point $\lambda_{1}=2$, $\lambda_{2}=-1$ ($z=2$). Two white lines are the energy gaps $\Delta_{gap}$ at different $d$.}
\label{fig:QC TO RC}
\end{figure}

\begin{figure}[htbp]
\includegraphics[scale=0.335]{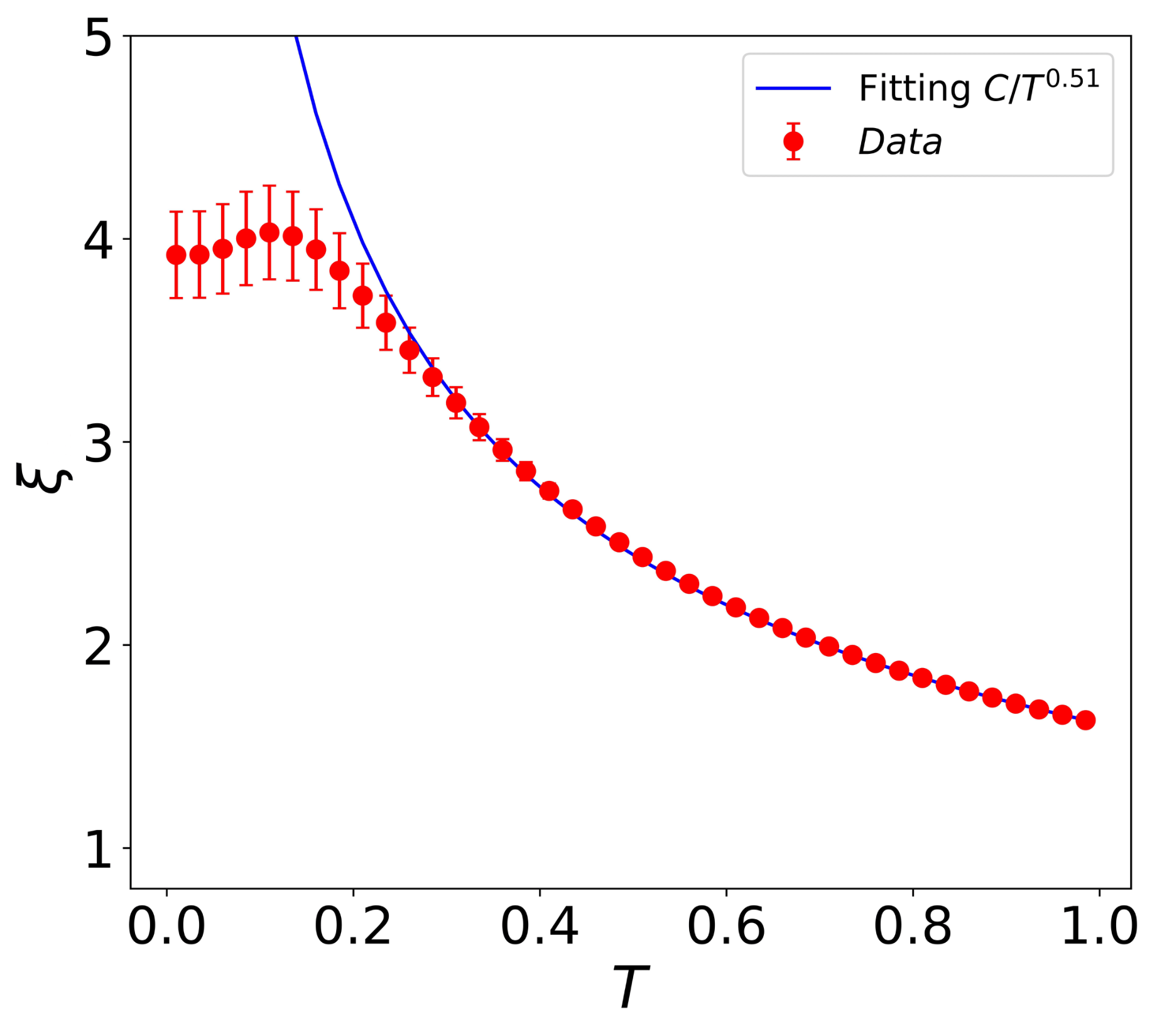}
\caption{Quantum Critical to Quantum Disordered. The correlation length $\xi$ vs temperature $T$ at $\lambda_{1}=1.8$,$\lambda_{2}=-1.2$ ($\Delta_{gap}\sim0.22$). $C\sim1.69$}
\label{fig:QC TO QD slice}
\end{figure}

\begin{figure}[htbp]
\begin{center}
\includegraphics[scale=0.335]{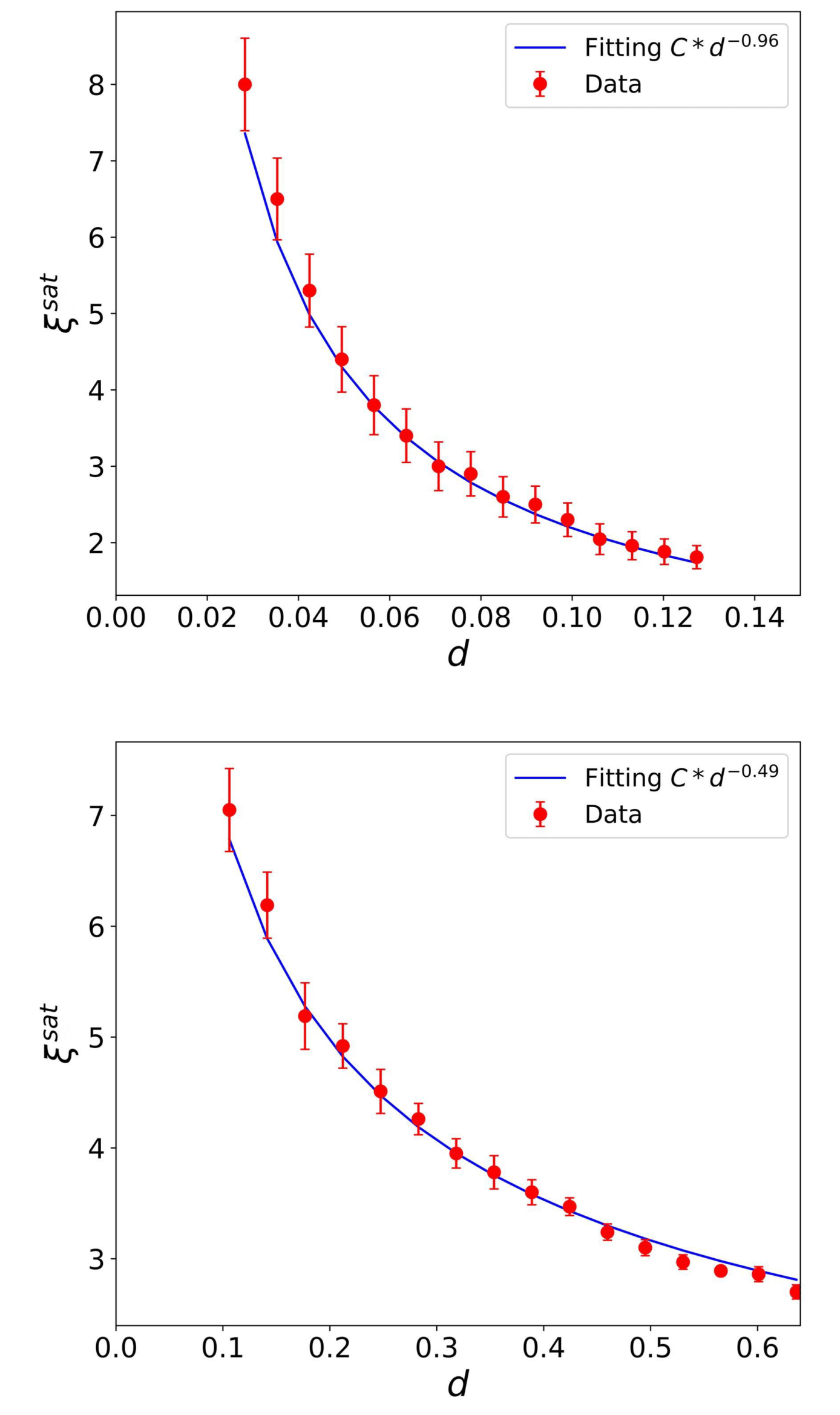}
\caption{The saturated correlation length $\xi^{sat}$ vs distance $d$ in quantum disordered regime. Top: $d$ is the distance to the multicritical point $\lambda_{1}=1$,$\lambda_{2}=0$ ($z=1$). Bottom: $d$ is the distance to the multicritical point $\lambda_{1}=2$,$\lambda_{2}=-1$ ($z=2$). $C_{top}\sim0.25$. $C_{bottom}\sim2.34$ }
\label{fig:sat cl}
\end{center}
\end{figure}

\begin{figure}[htbp]
\begin{center}
\includegraphics[scale=0.335]{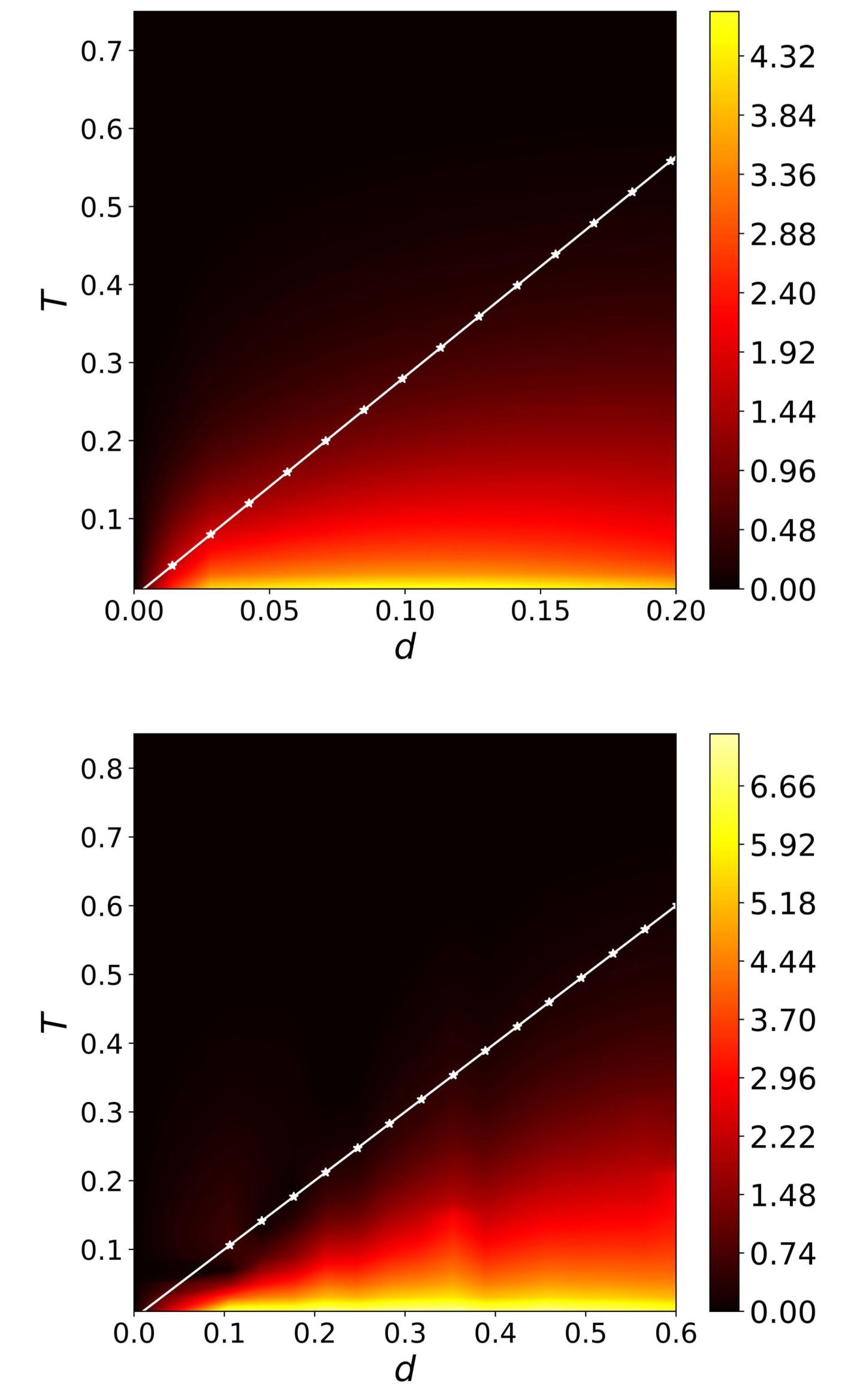}
\caption{Quantum critical fan diagrams. Black: Quantum Critical. Yellow: Quantum Disordered. Top: $d$ is the distance to the multicritical point $\lambda_{1}=1$,$\lambda_{2}=0$ ($z=1$). Bottom: $d$ is the distance to the multicritical point $\lambda_{1}=2$,$\lambda_{2}=-1$ ($z=2$).}
\label{fig:QC TO QD}
\end{center}
\end{figure}

\subsection{Quantum Critical to Renormalized Classical} 
Full construction of quantum critical fans along the three critical lines is not necessary since the quantum critical fan diagrams are quantitatively the same.  Here we choose to construct the quantum fan diagrams at two points $\lambda_{1}=2$, $\lambda_{2}=-1$ along the line $\lambda_{2} = \lambda_{1} - 3$ and $\lambda_{1}=1$, $\lambda_{2}=0$ along the line $\lambda_{2} = \lambda_{1} - 1$. Referring to Fig.~\ref{fig:critical}, at two critical points, we see the correlation length $\xi$ scale as $1/T^{1/z}$ with $z$ the theoretical values $1$($2$). Moving away from the critical point ($\lambda_{1}=2$, $\lambda_{2}=-1$) into the ordered phase along the line $\lambda_{2}=\lambda_{1}-3$, we see a crossover happens when we plot $\xi$ vs $T$ in Fig.~\ref{fig:QC TO RC slice}. At high $T$ ($T>\Delta_{gap}$), the temperature dependence of the correlation length is characterized by a power law.  At low $T$ ($T<\Delta_{gap}$), the correlation length grows exponentially to infinity. Between these regimes, we have a crossover. From that, we construct quantum critical fan diagrams along the lines $\lambda_{2} = \lambda_{1} - 1$ and $\lambda_{2} = \lambda_{1} - 3$. These are shown in Fig.~\ref{fig:QC TO RC}. The colors in these  plots represent the relative deviation from the power law scaling. The white lines are obtained by computing the energy gap at different $d$. Since we are considering crossovers, not phase transitions, we do accept some misalignments between the places where the color change happens and the white lines.  

\subsection{Quantum Critical to Quantum Disordered} 
 Moving away from the critical point ($\lambda_{1}=2$,$\lambda_{1}=-1$) into the disordered phase, we also see a crossover happens when we plot $\xi$ vs $T$ in Fig.~\ref{fig:QC TO QD slice}. At high $T$ ($T>\Delta_{gap}$), the temperature dependence of the correlation length is again a power law. Interestingly, we see a bump in the intermediate region ($T\sim \Delta_{gap}$). We can not differentiate whether the occurrence of the bump is due to the model itself or uncertainty from the fitting. At low $T$ ($T<\Delta_{gap}$), we see a completely different behavior. The correlation length saturates to a finite value. We denote this value as $\xi^{sat}$, which depends on the distance $d$ to the critical point. 
\begin{equation}
\xi^{sat}\sim d^{-\nu}
\end{equation}
where $\nu$ is the critical exponent. It is clear from Fig.~\ref{fig:sat cl} that $\nu$ are close to 1 ($z=1$) and 1/2 ($z=2$).
Similarly, we construct the other parts of the quantum critical fan diagrams along the lines $\lambda_{2} = \lambda_{1} - 1$ and $\lambda_{2} = \lambda_{1} - 3$. These are shown in Fig.~\ref{fig:QC TO QD}. The colors in these contour plots again represent the relative deviations from the power law scaling. 

\section{Summary and Discussion}
In this paper, we have discussed several properties of an exactly solved model that exhibits three interesting quantum critical lines and two multi-critical points. Two multicritical points have different dynamical critical exponents $z$. The three critical lines have their own unique characteristics. On one line, the criticality is located at $k=\pm \pi$. The other line has its criticality located at $k=0$. The criticality on the third line is located at incommensurate $k$ points. At finite temperatures, quantum critical fans are built upon these critical lines so the phase diagram splits into three regimes (quantum critical, quantum disordered and renormalized classical) The correlation length $\xi$ obtained from the calculation in each regime has its special behavior on temperature. In quantum critical regime, $\xi$ scales as $1/T^{1/z}$ with $z$ depends on the critical point. In quantum disordered regime, $\xi$ becomes temperature independent. But the saturated value of $\xi^{sat}$ scales as $d^{-\nu}$. Both $z$ and $\nu$ determine the size of the critical fan.
In renormalized classical regime, $\xi$ grows as an exponential function in terms of $T$ as we approach  zero temperature. Finally, we construct the quantum critical fan along two different lines.  

In the future, one could add further neighbor interaction while still maintaining the integrability of the model. This will make fine-tuning a lot easier. Then one could explore the quantum critical fan from a critical surface. However, such a model may be difficult to be realized in experiments. 
\section*{ACKNOWLEDGMENTS}
H.Y. was supported by M.L Bhaumik Institute for Theoretical Physics at UCLA. S.C. was supported by funds from the David  S. Saxon Presidential Term Chair.

%

\end{document}